\documentclass{PoS}

\usepackage{graphicx}
\usepackage{epsfig}
\usepackage{subfigure}
\usepackage{latexsym}
\usepackage{amsmath}
\usepackage{float}

\newcommand{\myvec}[1]{{\bf #1}} %{\mbox{\boldmath{$#1$}}}
\newcommand{\be}{\begin{equation}}
\newcommand{\ee}{\end{equation}}
\newcommand{\bea}{\begin{eqnarray}} % only untightened
\newcommand{\eea}{\end{eqnarray}}

\renewcommand{\i}{{\rm{i}}}
\newcommand{\SU}[2]{\ensuremath{{\rm{SU}}_{\rm{ #1 }}( #2 )}}

\title{Schr\"odinger functional boundary conditions and improvement of the $\SU{}{N}$ pure gauge action for $N>3$}

\ShortTitle{SF boundary conditions and $\mathcal{O}(a)$ improvement}

\author{\speaker{Tuomas Karavirta}\\
        {CP}$^{ \bf 3}${-Origins},
IFK \& IMADA, University of Southern Denmark,\\ 
Campusvej 55, DK-5230 Odense M, Denmark.\\
        E-mail: \email{karavirta@cp3-origins.net}}
        
        \author{Ari Hietanen\\
       {CP}$^{ \bf 3}${-Origins},
IFK \& IMADA, University of Southern Denmark,\\ 
Campusvej 55, DK-5230 Odense M, Denmark.\\
        E-mail: \email{hietanen@cp3-origins.net}}
        
\author{Pol Vilaseca\\
        School of Mathematics, Trinity College Dublin,\\
        Dublin 2, Ireland\\
        E-mail: \email{dionaea.0@gmail.com}}

\abstract{The leading method to study the running coupling constant of
  non-abelian gauge theories is based on the Schr\"odinger functional
  scheme. However, the boundary conditions and $\mathcal{O}(a)$ improvement have not  been systematically generalized for theories with more than three
  colors. These theories have applications in BSM model building as
  well as in the large $N$ limit. We have studied the boundary conditions
  and improvement for the pure Yang-Mills theory within the Schr\"odinger
  functional scheme. We have determined for all values of $N$ the
  boundary fields which provide high signal/noise ratio. Additionally,
  we have calculated the improvement coefficient $c_t$ for the pure
  gauge to one loop order for $\SU{}{N}$ gauge theories with
  $N=2,\ldots,8$ from which $N\geq 4$ are previously unknown. \\ \\ \\ $\textit{Preprint: CP$^3$-Origins-2013-042 DNRF90 \& DIAS-2013-42}$}

\FullConference{31st International Symposium on Lattice Field Theory - LATTICE 2013\\
		July 29 - August 3, 2013\\
		Mainz, Germany}

\begin{document}

\section{Introduction}
Recently there has been an interest in the scaling properties of the
gauge coupling in $\SU{}{N}$ theories with more than three colors
\cite{Lucini:2008vi,DeGrand:2012qa,Hietanen:2012ma}. For a review of large $N$ gauge theories see \cite{Lucini:2012gg}. The studies are
motivated by their applications in Beyond Standard Model (BSM) physics and by the
need to understand the scaling of the coupling constant in the large
$N$ limit. 

The main tool for measuring the evolution of the coupling constant as
a function of the scale on the lattice is the Schr\"odinger functional
method. As is well know this method suffers from $\mathcal{O}(a)$
lattice artifacts that originate from the boundary terms. 
They can be
removed by adding improvement terms to the
action and tuning appropriately the improvement coefficients \cite{Luscher:1992an}.  
Until now the boundary  
coefficient necessary to improve the pure gauge theory was only known for $N=2$ and $3$ \cite{Luscher:1992an,Luscher:1993gh}.  

In these proceedings, we present the preliminary results of calculating
the improvement coefficient $c_t$ to one loop order in perturbation
theory for the pure gauge theory with $N=2,\ldots,8$. 
After implementing $\mathcal{O}(a)$ improvement 
we see that the discretization effects are reduced for all considered 
values of $N$. 
In addition,  we
have done a study with a general $N$ to find Schr\"odinger functional
boundary fields with high signal/noise ratio that could be used in
lattice simulations.   

\section{Theory}
We use the standard $\mathcal{O}(a)$ improved $\SU{}{N}$ Wilson gauge action in the Schr\"odinger functional scheme
\begin{equation}
S=S_G+\delta S_{G,b},
\end{equation}
where
\bea
S_G&=&\frac{1}{g_0^2}\sum_{p}{\rm{Tr}}[1-U(p)],\\
\delta S_{G,b}&=&\frac{1}{g_0^2}({c_t}-1)\sum_{p_t}{\rm{Tr}}[1-U(p_t)].
\label{deltaSGb}
\eea
$U(p_t)$ refer to the timelike plaquettes on the $T=0$ and
$T=L$ boundaries. Additionally the perturbative expansion of the
improvement coefficient $c_t$ is 
\be
{c_t}=1+c_t ^{(1,0)} g_0 ^2+\mathcal{O}(g_0 ^4).
\ee
The gauge fixing procedure leads to the addition of gauge fixing
$S_{gf}$ and Faddeev-Popov ghost $S_{FP}$ terms to the action.

In the Schr\"odinger functional (SF) scheme the boundary conditions in the temporal boundaries are taken to be 
abelian and spatially constant \cite{Luscher:1992an}, given by
\begin{equation}
U_k(t=0,\vec{x})=\exp[a C_k],\quad U_k(t=L,\vec{x})=\exp[a C'_k],
\nonumber
\end{equation}
with
\begin{equation}
C_k=\frac{\i}{L}{\rm{diag}}(\phi_1({\eta}),\dots,\phi_n({\eta})),\quad C'_k=\frac{\i}{L}{\rm{diag}}(\phi'_1({\eta}),\dots,\phi'_n({\eta})).
\nonumber
\end{equation}
The phases $\phi_{i}$ and $\phi_{i}'$ depend on an internal parameter $\eta$.
This choice of boundary conditions induces a background field which
is a unique minimum of the action provided that the phases $\phi$ and
$\phi'$ lay within the fundamental domain
\cite{Luscher:1992an}. See Section~\ref{sec:boundary} for more details.
Boundary conditions in the spatial 
directions are taken to be periodic.

%In the Schr\"odinger Functional (SF) scheme we introduce a constant background 
%field by setting boundary conditions for the gauge fields at times
%$T=0$ and $T=L$. The boundary conditions are periodic in to the
%spacial directions. The   boundary conditions introduce
%${\mathcal{O}}(a)$ contributions to the action, which can be canceled
%by the boundary counter term  
%$\delta S_{G,b}$ in the action \ref{deltaSGb}. The boundary conditions
%in the Schr\"odinger functional scheme can be written as 
%\begin{equation}
%U_k(t=0,\vec{x})=\exp[a C_k],\quad U_k(t=L,\vec{x})=\exp[a C'_k],
%\nonumber
%\end{equation}
%where
%\begin{equation}
%C_k=\frac{\i}{L}{\rm{diag}}(\phi_1({\eta}),\dots,\phi_n({\eta})),\quad C'_k=\frac{\i}{L}{\rm{diag}}(\phi'_1({\eta}),\dots,\phi'_n({\eta})).
%\nonumber
%\end{equation}
%The system has a unique minimal action if the fields $\phi$ and
%$\phi'$ are elements of the fundamental domain
%\cite{Luscher:1992an}. See Section~\ref{sec:boundary} for more details.

The effective action has a perturbative expansion of the form\footnote{We refer to the original literature \cite{Luscher:1992an} for details
on the perturbative expansion of the effective action.}
\begin{equation}
\Gamma=-\ln\left\{ \int D[\psi] D[\bar{\psi}] D[U] D[c] D[\bar{c}] e^{-S}\right\}=g_0 ^{-2}\Gamma_0+\Gamma_1+\mathcal{O}(g_0 ^2).
\label{efac}
\end{equation}
A renormalized coupling can be defined in
the Schr\"odinger functional scheme as
a derivative of the effective action \eqref{efac} respect to the parameter 
$\eta$,
\begin{equation}
g^2=\frac{\partial \Gamma_0}{\partial{\eta}} / \frac{\partial \Gamma}{\partial{\eta}}=g_0 ^2-g_0 ^4 \frac{\partial \Gamma_1}{\partial{\eta}} / \frac{\partial \Gamma_0}{\partial{\eta}} +\mathcal{O}(g_0 ^6). 
\label{p1}
\end{equation}

In lattice studies it is common to use the lattice step scaling
function $\Sigma(u,s,L/a)$, which describes the evolution of the renormalized coupling
constant under a change of scale by a factor $s$:
\begin{eqnarray}
\Sigma(u,s,L/a)&=&g^2(g_0,sL/a)\vert_{g^2(g_0,L/a)=u}\nonumber\\
&=&u+\Sigma_{1,0}(s,L/a) u^2+\mathcal{O}(u^3).
\end{eqnarray}
We also use the function 
\begin{equation}
\delta_0(L/a)=\frac{\Sigma_{1,0}(2,L/a)}{\sigma_{1,0}(2)},
\end{equation}
to measure the convergence of the lattice step scaling function $\Sigma_{1,0}(2,L/a)$ to its continuum limit $\sigma_{1,0}(2)=2b_{0,0}\ln 2$. In the previous equation $b_{0,0}=11N_c/(48\pi^2)$ is the one loop coefficient of the pure gauge $\beta$-function.

\section{Boundary fields \label{sec:boundary}}
\begin{figure}[h]
\centering
\includegraphics[scale=1.2]{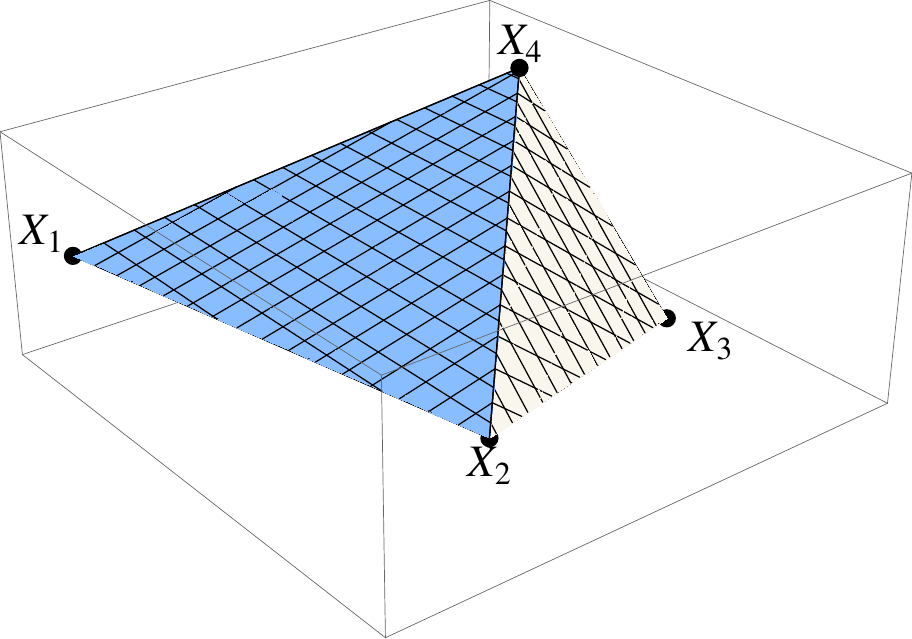}
\caption{Fundamental domain of $\SU{}{4}$}
\label{FD SU4}
\end{figure}

Boundary fields $\phi$ and $\phi'$ are within the fundamental domain if they satisfy the equations  \cite{Luscher:1992an}
\begin{equation}
\phi_1< \phi_2 < \ldots <\phi_n,\quad |\phi_{i}-\phi_{j}|<2\pi,\quad \sum_{i=1} ^N \phi_i=0.
\end{equation}
Such vectors $\phi$ form a $N-1$ simplex with vertices
\be
\begin{array}{lcl}
\myvec{X}_1 &= &\frac{2\pi}{N}\left(-N+1,1,1,\dots,1\right)\\
\myvec{X}_2 &= &\frac{2\pi}{N}\left(-N+2,-N+2,2,\dots,2\right)\\
\myvec{X}_3 &= &\frac{2\pi}{N}\left(-N+3,-N+3,-N+3,3,\dots,3,\right) \\
& \vdots & \\
\myvec{X}_{N-1} &= & \frac{2\pi}{N}\left(-1,-1,\dots,-1,N-1\right)\\
\myvec{X}_N &= & (0,0,\dots,0).
\end{array}
\ee
For $\SU{}{4}$ the fundamental domain is shown in figure \ref{FD SU4}.

\begin{figure}[h]
\centering
\includegraphics[scale=1.2]{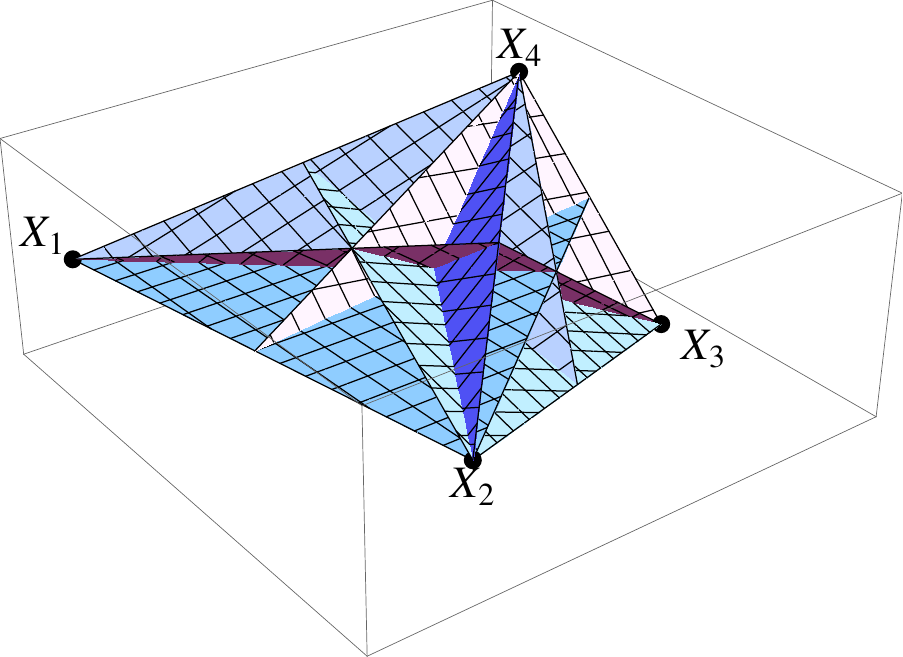}
\caption{All possible $R_{i,j}(\phi)$ hyperplanes on the fundamental domain of $\SU{}{4}$}
\label{Hyperplane}
\end{figure}

We start from the conjecture that the signal to noise ratio is maximized if $\phi$
and $\phi'$ are chosen, s.t. they are as far from the edges of the
fundamental domain and each other as possible \cite{Lucini:2008vi,Luscher:1992an,Luscher:1993gh}. 
To find such points we
first want to define a mapping which mirrors the point in the
fundamental domain. We start by defining a mapping
$R_{i,j}(\phi)$ that reflects the points in the
fundamental domain with respect to a $(N-2)$ dimensional
hyperplane. The hyperplane $R_{i,j}(\phi)$ goes through vertices
$\myvec{X}_k$, $k\neq i,j$  and intersects the line connecting
$\myvec{X}_i$ and $\myvec{X}_j$ at the middle. In figure
\ref{Hyperplane} we show all the possible mappings $R_{i,j}(\phi)$ on
$\SU{}{4}$.  

The function $R_{i,j}(\phi)$ is not a mapping from the fundamental
domain to itself, but we can define a composite mapping\footnote{$R_{i,i}(\phi)$ is the identity mapping and
  $[x]$ means the integer part of $x$}
\begin{equation}
M(\phi)=\left(R_{1,N-1}\circ R_{2,N-2}\circ\ldots\circ
R_{[N/2],N-[N/2]}\right)(\phi),
\label{M}
\end{equation}
 that has this property. The value of the field $\phi'$ is then derived from $\phi$ using $M(\phi)$ which has a simple form $\phi'_i=\phi_{N-i+1}$.

We choose $\phi$ to be in the middle of a line connecting $\myvec{X}_1$ and the centeroid of the fundamental domain and associate a flow\footnote{The normalization $\frac{\eta N}{2\pi(N-2)}$ is chosen such that the coefficients of $\eta$ the standard case of $\SU{}{3}$.}
\begin{eqnarray}
  t(\eta) & =& \frac{\eta N}{2\pi(N-2)} \left(\myvec{X}_1-\myvec{X}_{N-1}\right)\nonumber\\
  & =& \left(-\eta,\frac{2\eta}{N-2},\dots,\frac{2\eta}{N-2},-\eta\right),
\end{eqnarray}
to the direction which gets mirrored by $R_{1,N-1}(\phi)$
transformation and points outside from the fundamental domain. As an
example boundary fields of $\SU{}{4}$ are
\begin{equation}\begin{array}{ll} 
    \phi=\left\{\begin{array}{c} -\eta - 9\pi/8, \\ \eta + \pi/8, \\  \eta + 3\pi/8, \\ -\eta + 5\pi/8, \end{array} \right. &
    \phi' =\left\{\begin{array}{c}  \eta - 5\pi/8, \\ -\eta - 3\pi/8, \\ -\eta - \pi/8, \\  \eta + 9\pi/8. \end{array} \right. 
    \end{array}
\end{equation}
Note that these boundary fields are different than those used in \cite{Lucini:2008vi}. The possible improvement in the signal/noise ratio should be determined with lattice simulations.

\section{Boundary improvement}
Improvement coefficient $c_t$ is previously know for $N=2,3$ to one 
loop order in perturbation theory. These values have been calculated 
by L\"usher et. al. in \cite{Luscher:1992an} for $\SU{}{2}$ and 
in \cite{Luscher:1993gh} for $\SU{}{3}$ resulting in 
$c_t ^{(1,0)}(\SU{}{2})=-0.0543(5)$ and $c_t ^{(1,0)}(\SU{}{3})=-0.08900(5)$. 
The method used in \cite{Luscher:1992an} and \cite{Luscher:1993gh} 
is also applicable to $N>3$ with some modifications. 

The details of the calculation of  $c_t ^{(1,0)}$ will be given in
\cite{sfpaper}. The calculation goes along the lines of
\cite{Luscher:1992an}. The idea of the process is to calculate
$p_{1,0}(L/a)=\frac{\partial \Gamma_1(L/a)}{\partial{\eta}} /
\frac{\partial \Gamma_0(L/a)}{\partial{\eta}}$ from \eqref{p1} as a
function of the lattice size $L/a$. This is done by solving a second
order difference relation to several different operators. In this way
we are able to solve $p_{1,0}(L/a)$ and consequently the running
coupling $g^2$ to one loop order in perturbation theory for a range in
$L/a\in\{6,8,10,\ldots,64\}$. 

The variable $p_{1,0}(L/a)$ has an 
asymptotic expansion in $L/a$ \cite{Luscher:1992an}
\be
p_{1,0}(L/a)\sim\sum_{n=0} ^{\infty} (r_n + s_n \ln(L/a))\left(\frac{a}{L}\right)^{n},
\label{eq:p10series}
\ee
where $s_0=2 b_{0,0}$ and $s_1=0$. The coefficient $c_t ^{(1,0)}$ 
is determined by demanding linear cutoff effects to be absent in \eqref{eq:p10series},
i.e. $r_1=2 c_t ^{(1,0)}$.
The problem is then to extract the coefficient $r_1$ as accurately as possible from the $p_{1,0}(L/a)$ data. To do this we used the "Blocking" method described in \cite{Luscher:1985wf}. Our preliminary results are shown in table \ref{results}.

\begin{table}
\center
\begin{tabular}{|c|c|c|}
\hline
$N$ & $c_t ^{(1,0)}$ & $\delta c_t ^{(1,0)}$  \\
\hline
2 & -0.0543 & 0.0002 \\
3 & -0.088 & 0.005 \\ 
4 & -0.1220 & 0.0002 \\
5 & -0.154 & 0.004 \\
6 & -0.1859 & 0.0008 \\
7 & -0.218 & 0.004 \\
8 & -0.249 & 0.004 \\
\hline
\end{tabular}
\caption{The values of $c_t ^{(1,0)}$ and estimated errors $\delta c_t ^{(1,0)}$ for $N=2,\ldots,8$.}
\label{results}
\end{table}

\begin{figure}[h]
\centering
\includegraphics[scale=0.43]{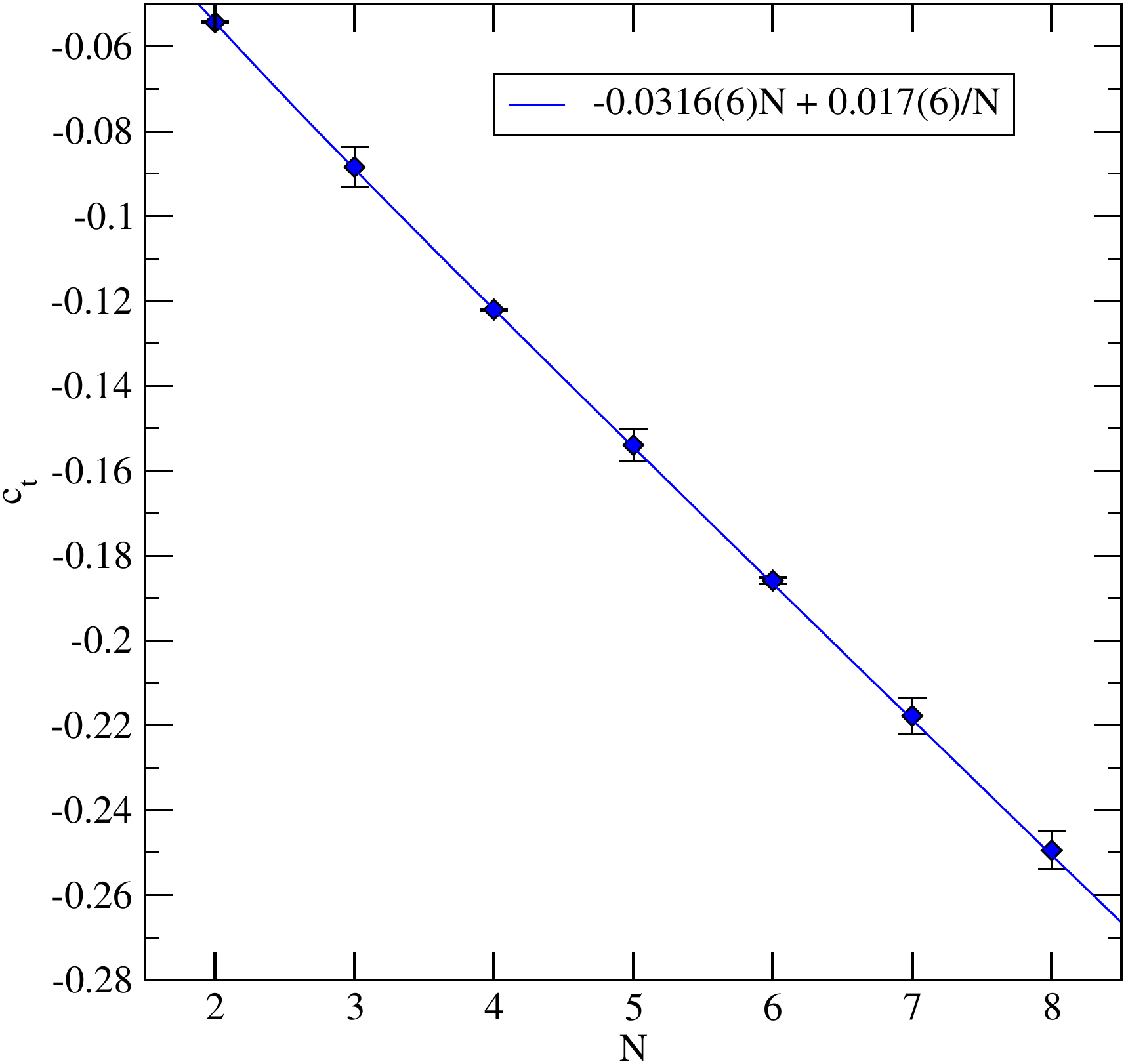}
\caption{$c_t ^{(1,0)}$ as a function of $\tilde{A} N+\tilde{B}/N$ fit to the data}
\label{ctfig}
\end{figure}

We expect $c_t ^{(1,0)}=A C_2(R)+B C_2(G)=\tilde{A}
N+\tilde{B}/N$. This is motivated by the fact that the Feynman
diagrams involved are proportional to these Casimir invariants. Also
it has been shown in  \cite{Karavirta:2011mv} that the fermionic part
of $c_t ^{(1)}$ is proportional to the Casimir invariant $T(R)$. See
also \cite{Sint:2012ae}. A plot of the values of $c_t ^{(1,0)}$ as a
function of $N$ and our fit to the data is shown in \ref{ctfig}. 

\begin{figure}[h]
\centering
\includegraphics[scale=0.63]{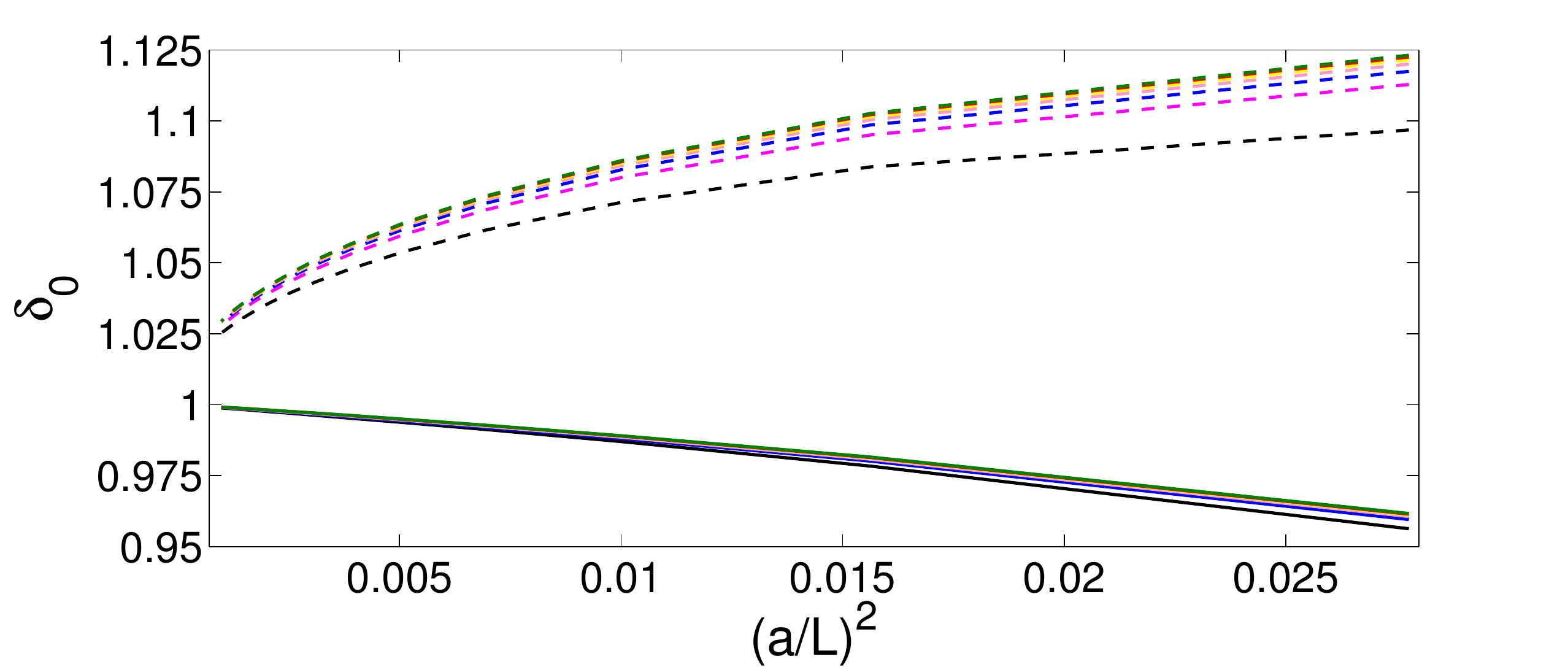}
\caption{The unimproved (dashed) and improved (solid) one loop lattice step scaling function normalized to the continuum limit ($\delta_0$) as a function of $(a/L)^2$ for $\SU{}{N}$ pure gauge with $N=2$ (black), $3$ (purple), $4$ (blue), $5$ (pink), $6$ (yellow), $7$ (red) and $8$ (green).}
\label{Stepscaling}
\end{figure}

We also want to be sure that setting $c_t$ to the value that we
obtained removes the $\mathcal{O}(a)$ terms form the lattice step
scaling function. This can be seen from figure \ref{Stepscaling} where we have plotted $\delta_0$ as a function of $(a/L)^2$. After the improvement $\delta_0$ behaves linearly which is a clear indication that the leading terms are of the order $\mathcal{O}(a^2)$.

\section{Summary and outlook}

We have investigated the boundary conditions in general $N$ and calculated the
$\mathcal{O}(a)$  boundary improvement coefficients 
for $N=2,\ldots,8$ in the Schr\"odinger
functional scheme. The precision in the determination of $c_t ^{(1,0)}$
can be increased by using more than double
precision floating point numbers in the numerical calculations. 
We are currently implementing these enhancements to improve our results.

\acknowledgments
This work was supported by the Danish National Research Foundation DNRF:90 grant. TK is also funded by the Danish Institute for Advanced Study.

%Our method suffers from lack of precision because we used double precision in deriving the perturbative values of the running coupling. To reliably extract the $c_t ^{(1,0)}$ coefficient from the data one needs higher accuracy. Also there seems to be better methods for extracting the  $c_t$ coefficient than the "Blocking" method that we used. One possibility would be to use the method described in \cite{Bode:1999sm}. We are currently working to resolve these issues by implementing a new numerical program that is capable of higher than double precision and testing several methods to extract the $c_t$ coefficient from our new data.

%\acknowledgments

\end{document}